\documentclass[aps,prx,showpacs,twocolumn,groupedaddress,superscriptaddress]{revtex4-1}

\usepackage{times,xspace}
\usepackage{graphicx,graphics,color,epsfig}
\usepackage{amsmath}
\usepackage{amssymb}
\usepackage{amsfonts}
\usepackage{amsfonts}
\usepackage{epstopdf}
\usepackage{bm}
\usepackage{hyperref}
\usepackage{color}
\usepackage{float}
\usepackage[normalem]{ulem}

\def\be{\begin{equation}}
\def\ee{\end{equation}}
\def\ba{\begin{eqnarray}}
\def\ea{\end{eqnarray}}

\newcommand{\figref}[1]{Fig.~\ref{#1}}

\begin{document}

\title{Role of phonon coupling in driving photo-excited Mott insulators towards a transient superconducting steady state}

\author{Sujay Ray}
\affiliation{Department of Physics, University of Fribourg, Fribourg-1700, Switzerland}
\author{Martin Eckstein}
\affiliation{Institute of Theoretical Physics, University of Hamburg, 20355 Hamburg, Germany}
\affiliation{The Hamburg Centre for Ultrafast Imaging, Hamburg, Germany}
\author{Philipp Werner}
\affiliation{Department of Physics, University of Fribourg, Fribourg-1700, Switzerland}

\date{\today}
 
\begin{abstract}
Understanding light-induced hidden orders is relevant for nonequilibrium materials control and future ultrafast technologies. Hidden superconducting order, in particular, has been a focus of recent experimental and theoretical efforts. In this study, we investigate the stability of light-induced $\eta$ pairing. Using a memory truncated implementation of nonequilibrium dynamical mean field theory (DMFT) and entropy cooling techniques, we study the long-time dynamics of the photoinduced superconducting state. In the presence of coupling to a cold phonon bath, the photodoped system reaches a quasi-steady state, which can be sustained over a long period of time in large-gap Mott insulators. We show that this long-lived prethermalized state is well described by the nonequilibrium steady state implementation of DMFT. 
\end{abstract}
\vspace{0.5in}

\maketitle

\section{Introduction}
Light-induced superconductivity in strongly correlated materials is an exciting research field stimulated by experimental discoveries of superconducting-like states \cite{Cavalleri2018, Fausti2011, phsc_FeSe, Buzzi_2020,phind_sc,phind_sc1}. The goal is to achieve superconductivity at room temperature in out-of-equilibrium systems with the help of ultra-short laser pulses \cite{Cavalleri2018}. Among the variety of strongly correlated systems, Mott insulators are particularly interesting, because the photodoped non-equilibrium states in large-gap Mott insulators can be long-lived and heating effects are suppressed  \cite{Murakami_rev,Strohmaier2010,Sensarma2010,Eckstein2011,Lenarcic2013,Lenarcic2014}. This provides a realistic posibility of stabilizing light-induced electronic orders on femtosecond or picosecond time-scales.

Numerous theoretical techniques like the density matrix renormalization group (DMRG), tensor network approaches and non-equilibrium dynamical mean field theory (DMFT) have been employed to simulate and understand photo-induced superconducting orders in Mott systems \cite{yuta_etasc,yuta_etasc1,eta_jiajun,eta_spintrip,eta_spintrip1,kaneko1,kaneko2}. These studies suggest that a staggered superconducting order ($\eta$-SC order) can be realized both in single-orbital and multi-orbital Hubbard models. Apart from these local spin-singlet and spin-triplet $\eta$-SC orders, the realization of chiral SC order with a $120^\circ$ phase twist in a triangular lattice system has also been discussed \cite{tri_chiral}. 

\begin{figure}[b]
\includegraphics[width=0.5\textwidth]{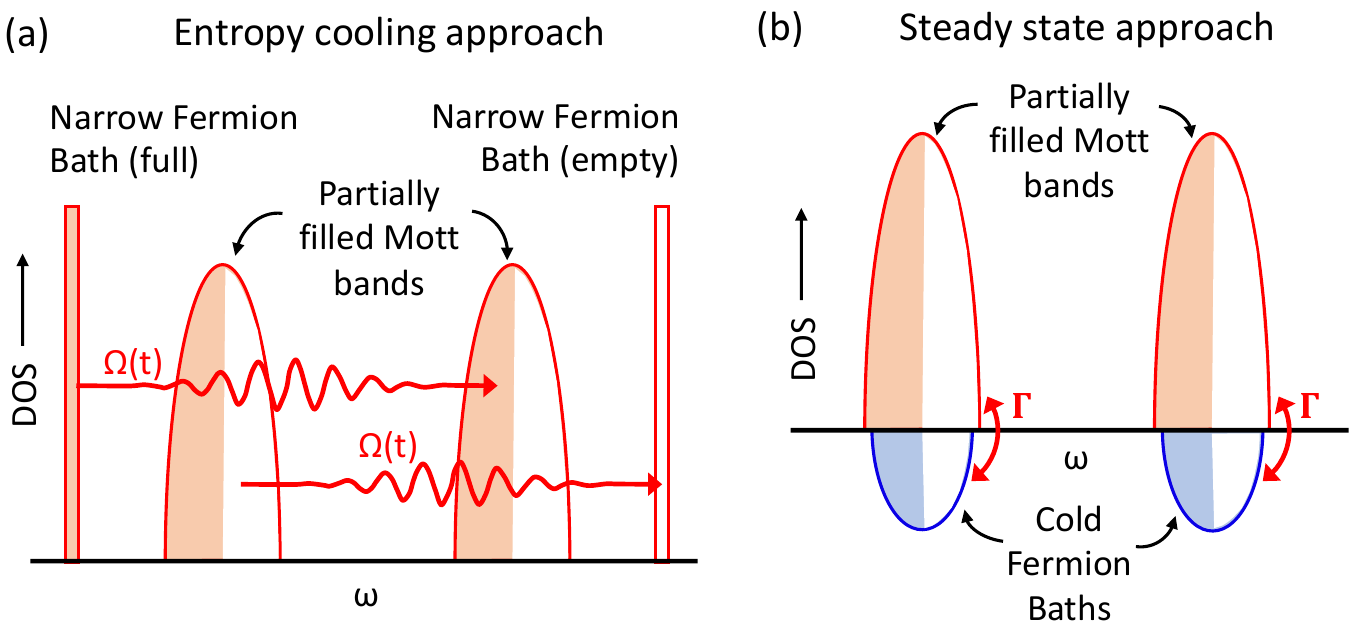}
\caption{Schematic illustrations of (a) the entropy cooling setup and (b) the non-equilibrium steady state (NESS) technique. In the entropy cooling case, the Hubbard bands are transiently coupled to full and empty Fermion baths with a narrow density of states. In the NESS approach, the Hubbard bands are permanently coupled to cold Fermion baths.}
\label{fig1}
\end{figure}

One of the primary challenges in stabilizing a staggered $\eta$-SC order or the $120^\circ$ condensate in photo-induced systems is to overcome the laser-induced heating. Real-time simulations based on nonequilibrium DMFT  \cite{NESSi} take into account the heating effect, as well as entropy reshufflings in the case of inter-band excitations. The latter effect is exploited in the entropy cooling approach \cite{entropy_cooling,entropy_cooling1} (see ~\figref{fig1} (a)), which has been used to reduce the heating effect during the preparation of a photo-doped state, leading to an effectively cold temperature and $\eta$-SC order in the case of an almost fully photo-doped Mott insulator \cite{entropy_cooling1}.

One important question which has not yet been studied in detail is the stability of the $\eta$-SC order after its induction by the laser pulse. Because of the large amount of energy injected during the laser excitation, the thermalized state in an isolated system will have a very high, or even negative temperature. We thus expect that the effective temperature of the photodoped state increases during the relaxation which follows after the pulse is switched off, and that the $\eta$-SC order decays. 
Several factors, like the gap size or photo-doping concentration, affect this dynamics. 

 One strategy to stabilize the $\eta$-SC order 
 is to introduce an energy dissipation mechanism which reduces the kinetic energy of the photo-excited charge carriers. In this study, we locally couple the electrons to cold phonon baths. In real materials, the specific heat of the lattice is much higher than that of the electrons, and the phonon dynamics is slow compared to the electronic dynamics, so that the phonon temperature can be assumed to remain low on sub-ps timescales. Such a phonon coupling prevents the system from heating up during the thermalization process and thus stabilizes the $\eta$-SC order even in systems with not too large Mott gaps.

While entropy cooling protocols and couplings to dissipative environments represent physical pathways for reaching a metastable phase, we may sometimes be less interested in this pathway than in the properties of the phase itself. In such a situation, the nonequilibrium steady-state (NESS) approach \cite{NESS_martin} illustrated in \figref{fig1}(b)) is more convenient. 
The NESS approach has been shown to essentially reproduce the photodoped Mott states obtained by explicit time propagation for small photodoping \cite{Mrtin_inchworm}, but the emergence of these quasi-steady states and their stability has not been studied in detail for electronically ordered phases. Here, we will fill this gap and show that the metastable $\eta$-SC state of photo-doped Mott systems is well described by the NESS approach.

The paper is structured as follows. In Sec.~\ref{sec_model}, we define the model and discuss the methods used in the real-time and NESS simulations. In Sec.~\ref{sec_results}, we present our findings -- the long-time dynamics of the $\eta$-SC state, the dependence of the relaxation dynamics on $U$, the effect of the $\eta$-SC order on the relaxation dynamics and the emergence of a long-lived prethermalized quasi-steady state in the system with and without a phonon coupling. Sec.~\ref{sec_conclusion} contains a short conclusion.

\section{Model and Method}
\label{sec_model}
We consider the single orbital repulsive Hubbard model on the Bethe lattice with coordination number $z$, described by the Hamiltonian
\begin{align}
&H = -t_\text{h}\sum_{\left<ij\right>,\sigma} c^{\dagger}_{i\sigma}c_{j\sigma} + U\sum_{i}n_{i\uparrow}n_{i\downarrow} - \mu \sum_{i}\left(n_{i\uparrow} + n_{i\downarrow} \right),
\label{eq_1}
\end{align}
 where $\sigma$ denotes the spin index, $n_{i\sigma}$ the spin-density at site $i$, $t_\text{h}$ is the nearest neighbor hopping amplitude between sites $i$ and $j$, $U$ is the Hubbard repulsion and $\mu$ the chemical potential. For large $U$, at half filling, this model has an antiferromagnetic ground state, while for high enough photo-doping and sufficiently low effective temperature, an $\eta$-SC state has been established by steady state DMFT calculations \cite{eta_jiajun}. To investigate the dynamics of the $\eta$-SC state, we employ real-time DMFT simulations and the entropy cooling technique \cite{entropy_cooling}. This approach has been previously used to realize an $\eta$-SC state in the extreme photo-doping limit (almost all sites either doubly occupied or empty) \cite{entropy_cooling1}. In the entropy cooling set-up, the Mott insulator is coupled to two non-interacting Fermion baths with narrow bandwidth (one completely full and the other completely empty) by a laser pulse, as schematically shown in ~\figref{fig1}(a). By applying a carefully tuned chirped pulse with a slowly increasing frequency, it is possible to realize photo-doped Mott insulators with a desired density of doublons (doubly occupied sites) and holons (empty sites). The reshuffling of entropy from the Mott insulator to the noninteracting baths furthermore helps to produce an effectively cold photo-doped system. Here, the effective temperature is defined by 
fitting a Fermi distribution function to the ratio of the occupied density of states $A^{<}(\omega)$ and total density of states $A(\omega)$ in the upper Hubbard band (or equivalently in the lower Hubbard band).
 
 Specifically, the photo-doping of the Mott insulator is achieved by modulating the hopping amplitude between the system and bath $v_{\text{sb}}$ as $v_{\text{sb}}(t) = A_0 f_{\text{en}}(t) \sin(\Omega(t)t)$, where $A_0$ denotes the amplitude of the pulse, $f_{\text{en}}(t)$ is an envelope function determining the switch-on and switch-off times of the hopping, and $\Omega(t)$ is a time-dependent (chirped) frequency. The DMFT self consistency condition, which relates the hybridization function $\Delta(t,t')$ of the DMFT impurity model to the Green's function $G(t,t')$ of the system \cite{Georges1996}, is implemented for the infinitely connected Bethe lattice ($z\rightarrow \infty$) and is given by
 \begin{align}
    \Delta(t,t^{\prime})=& \, v_{\text{h}}(t)\gamma G(t,t^{\prime})\gamma v_{\text{h}}(t^{\prime}) \nonumber\\
    &+ \sum_{\alpha}v_\text{sb}(t) G_{\text{bath},\alpha}(t,t^{\prime})v_\text{sb}(t'),
    \label{eq_2}
\end{align}
where the renormalized hopping $v_{\text{h}}(t)=v_{\text{h}}\propto t_\text{h}/\sqrt{z}$ is related to the bandwidth $W$ of the system by $W=4v_\text{h}$. $G_{\text{bath},\alpha}(t,t^{\prime})$ is the Green's function of the non-interacting Fermion bath with flavor $\alpha$ (full,empty). The hybridization function and Green's function are written using the Nambu spinor basis $\psi^{\dagger} = (c_{\uparrow}^{\dagger} \hspace{0.2cm} c_{\downarrow})$. In this basis, the $\gamma$ matrix is defined as the unity matrix for the purpose of studying $\eta$-SC order.

With the Hamiltonian defined in ~\eqref{eq_1} and 
the 
hybridization function defined in ~\eqref{eq_2}, after the laser pulse is switched off, the narrow Fermion baths in the entropy cooling set-up become decoupled. As a result, the $\eta$ pseudospin is conserved under the time evolution. To study a non-trivial dynamics of the $\eta$ order parameter, we add a next nearest neighbor hopping term $v_{\text{h}}^{\text{NN}}$. The self-consistency equation for the Bethe lattice without the narrow Fermion baths is then modified as
\begin{eqnarray}
\Delta(t,t^{\prime})=v_{\text{h}}(t)\gamma G(t,t^{\prime})\gamma v_{\text{h}}(t^{\prime}) + v_{\text{h}}^{\text{NN}}(t)\sigma_{z} G(t,t^{\prime})\sigma_{z}v_{\text{h}}^{\text{NN}}(t^{\prime}), \nonumber \\
    \label{eq_3}
\end{eqnarray}
where $\sigma_{z}$ is a Pauli spin matix. In the $\eta$-SC state, the SC order parameter or off-diagonal element of  $G(t,t^{\prime})$ changes sign between sublattice $A$ and sublattice $B$. This explains the $\gamma$ matrices in the first term, which represents hopping processes between sublattices, and the $\sigma_z$ matrices in the second term, which represents hopping processes within the same sublattice. In all our calculations, we use $v_{\text{h}}^{\text{NN}}/v_{\text{h}}=0.25$.  

The time evolution of the photo-doped system is simulated with a nonequilibrium DMFT framework based on NESSi \cite{NESSi}, which calculates the two-time Green's function $G(t,t^{\prime})$ by solving the Kadanoff-Baym equations. This method becomes costly for long-time simulations because of the increasing memory and CPU demand. Since we are interested in the long-time dynamics, where the system reaches a quasi-steady state, we need to employ a more efficient scheme based on memory truncation \cite{Schueler2018,trunc_NESSi}.
In this truncation scheme, the self energy is assumed to decay within a cutoff window $t_{\text{c}}$ on the real time axis (a suitable  $t_{\text{c}}$ can be chosen by observing the decay of the hybridization function). With such an approximation, the computational effort and memory cost for a given time step is controlled by $t_c$ rather than by the maximum simulation time $t_\text{max}$, allowing us to simulate the long-time dynamics in photo-doped Mott insulators. 

In this study, we want to explore how and under which conditions a Mott insulator reaches a quasi steady state after photo-excitation, and if this steady state is well reproduced by NESS simulations. In the latter approach \cite{NESS_martin}, the upper Hubbard band (UHB) and lower Hubbard band (LHB) are weakly coupled to cold Fermion baths, as shown in ~\figref{fig1}(b). These Fermion baths inject electrons in the UHB and remove electrons from the LHB and thus convert the half-filled Hubbard model into an effective photo-doped system. The NESS simulation is implemented for the infinitely connected 
Bethe lattice, where 
the self-consistency equation is given by 
\begin{align}
\Delta(t,t^{\prime})& = & v_\text{h}^{2}\gamma G(t,t^{\prime})\gamma + (v_{\text{h}}^{\text{NN}})^{2}\sigma_{z} G(t,t^{\prime})\sigma_{z} \nonumber \\
    &&+ \sum_{b=1,2} D_{\text{b}}(t,t^{\prime}).
    \label{eq_10}
\end{align}
Here $D_{\text{b}}$ represents the contributions of the Fermion baths, which are defined in frequency space as $D_{\text{b}}(\omega)=g^{2}\rho_{b}(\omega)=\Gamma\sqrt{W_b^2-(\omega-\omega_b)^2}$, where $\Gamma=g^2/W_b^2$ is a dimensionless coupling constant, $\omega_b$ is the center of the energy spectrum and $W_b$ is the half-bandwidth of the bath $b$. We use the NCA (non-crossing approximation) as impurity solver in both the real-time entropy cooling and steady state calculations \cite{Eckstein2010}. 

NESS simulations offer excellent control over the amount of photo-doping, which can be defined by the doublon density $d=\left<n_{\uparrow} n_{\downarrow} \right>$, and also a good level of control over the effective temperature, which can be defined by fitting a Fermi distribution function to the ratio of the occupied density of states and total density of states within the energy range of the UHB and LHB. By tuning the chemical potentials of the Fermion baths $\mu_b$, we can tune the density of photocarriers (doublons and holons) and by tuning the temperature of the Fermion baths $T_b$, we can adjust the effective temperature. In our calculations, we define the staggered $\eta$-SC order as
\begin{eqnarray}
    \eta^{x}= \delta_{i}^{A/B}\frac{1}{2}(c^{\dagger}_{\uparrow}c^{\dagger}_{\downarrow} + \text{H.c.}),
    \label{eq_11}
\end{eqnarray}
where $\delta_{i}^{A/B}$ is $+1$($-1$) on sublattice $A$($B$). In order to study the $\eta$-SC phase, we apply a small seed field $P_{\text{seed}}=0.001$, which is coupled to the $\eta^{x}$ order parameter. In a symmetry broken phase the value of the order parameter grows to a high value, independent of the value of the seed field.

In our calculations, we also study the effect of a local phonon coupling \cite{Martin_phonon}. For this, we attach a boson bath to each site through a Holstein coupling. The local Hamiltonian becomes
\begin{eqnarray}
H_{\text{loc}}&=&Un_{\uparrow}n_{\downarrow} + \omega_{0}\left(b^{\dagger}b+\frac{1}{2}\right) -\mu(n_{\uparrow}+n_{\downarrow}) \nonumber \\
&& 
+ g(b^{\dagger} + b)(n_{\uparrow}+n_{\downarrow}-1),
\label{eq_12}
\end{eqnarray}
where $b^{\dagger}$ ($b$) are bosonic creation (annihilation) operators, $\omega_{0}$ is the frequency of the phonon and $g$ is the phonon coupling strength. We stay within the weak coupling regime, where the Migdal-type electron-phonon diagrams leads to the self-energy  
\begin{eqnarray}
    \Sigma(t,t^{\prime})=\Sigma_{\text{loc}}(t,t^{\prime})+g^{2}G(t,t^{\prime})D(t,t^{\prime}).
    \label{eq_13}
\end{eqnarray}
Here $\Sigma(t,t^{\prime})$ is the total self-energy, $\Sigma_{\text{loc}}$ is the local electronic contribution and the last term is the electron-phonon self-energy with $D(t,t^{\prime})$ denoting the free boson propagator. The coupling strength $g$ and the frequency of the phonon $\omega_{0}$ are chosen such that the effective coupling $g^{2}/\omega_{0}<1$. In all our calculations we take $\omega_{0}=0.4$, which is much smaller than the Mott gap, and $g=0.2$. 

In the NESS simulations, the two-time contour functions 
depend only on the time difference and Eq.~\eqref{eq_13} is modified accordingly.

\section{results}
\label{sec_results}
\subsection{Dynamics of photodoped Mott insulators} 
\begin{figure}[t]
\includegraphics[width=0.48\textwidth]{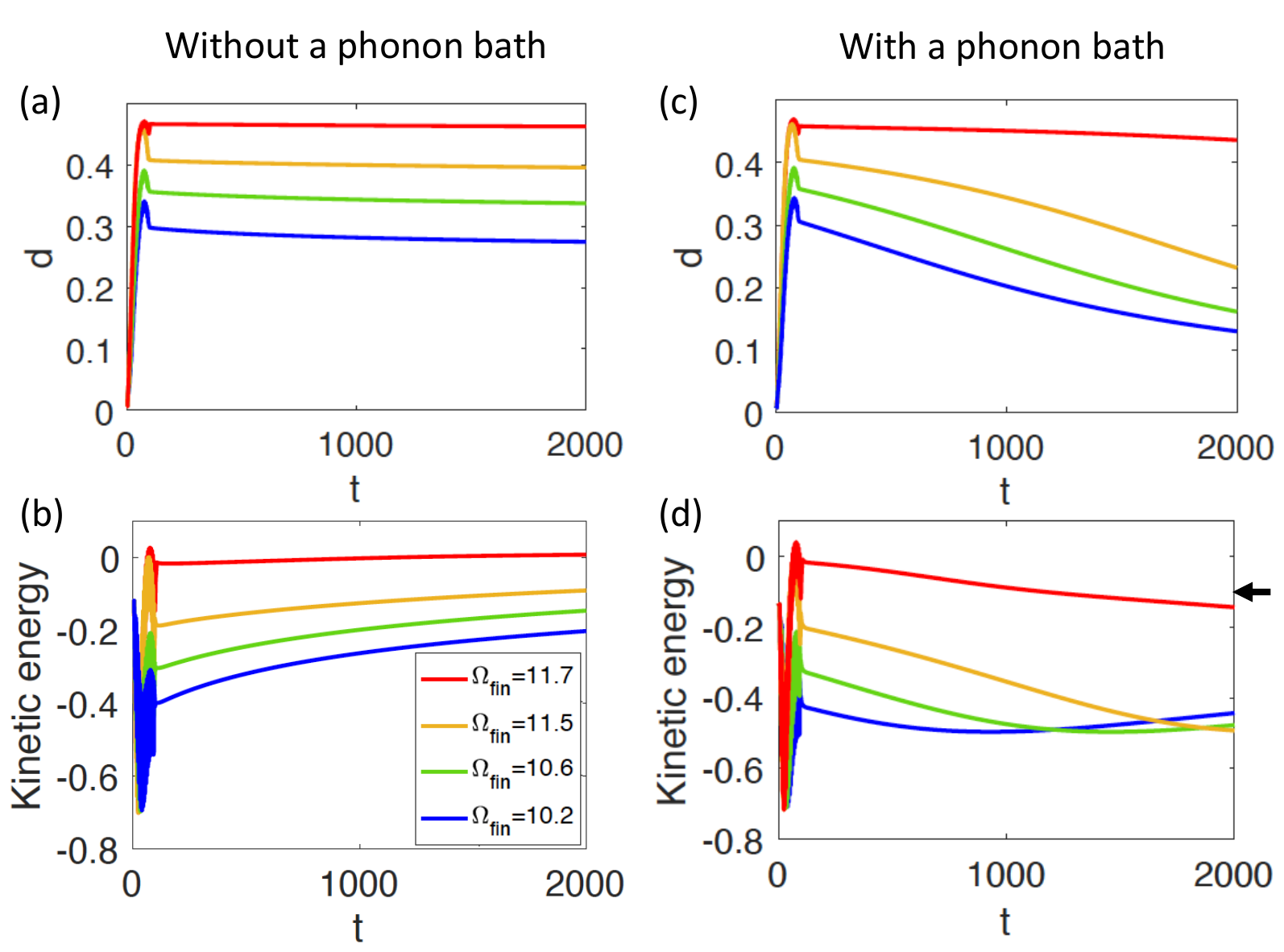}
\caption{(a) Doublon densities and (b) kinetic energies during and after chirped pulses with indicated $\Omega_\text{fin}$ in the system without phonon bath coupling. (c) Doublon densities and (d) kinetic energies as a function of time with a weakly coupled phonon bath. All the results are obtained for $U=9$ and different photodopings are reached with suitably chosen short chirped pulses. The frequencies at the end of the frequency ramp, $\Omega_{\text{fin}}$, are shown in (b). Same colored curves correspond to the same system in all the four figures.
The arrow in panel (d) shows the value of the kinetic energy in the fully thermalized state.
}
\label{fig2}
\end{figure}

We first study the dynamics of the photodoped single orbital Hubbard model using the entropy cooling technique and memory-truncated nonequilibrium DMFT simulations. In all our calculations we set $v_{\text{h}}=1$, which yields a bandwidth of approximately 4 for the UHB and LHB. We choose $U=9$ and use different frequency pulses to induce different levels of photo-doping. We use a chirped frequency $\Omega(t)=\Omega_{\text{ini}}+(\Omega_{\text{fin}}-\Omega_{\text{ini}})\sin(\pi t/2t_{\text{ramp}})$, where $\Omega_\text{fin}>\Omega_\text{in}$ and $t_{\text{ramp}}$ roughly corresponds to the duration of the pulse. The two narrow Fermion baths of bandwidth $0.05$ are placed at $\omega=\pm 6.0$, and we choose $\Omega_\text{ini}=7.5$. Figure~\ref{fig2} shows the time evolution of the doublon density and kinetic energy for the indicated $\Omega_\text{fin}$. Due to particle-hole symmetry, the holon density is the same as the doublon density. As one can deduce from panel (a) the pulses with larger $\Omega_\text{fin}$ produce a higher density of photo-carriers. In ~\figref{fig2}(a) we show the doublon density as a function of time without coupling the system to an external boson bath. Because of the lack of any mechanism for dissipating energy, the doublon density remains high after the pulse. In fact, keeping the total energy conserved, these systems will thermalize to a negative temperature state with double occupation $d>0.25$ and a positive kinetic energy. The heating of the photo-carriers during this thermalization process is evident from the kinetic energy of the system, which is plotted in ~\figref{fig2}(b). We observe an increase of the kinetic energy for all photodopings, which indicates that higher energy states get populated. We furthermore note that for lower photodoping, the rate of increase of the kinetic energy is faster. This is because for the chirped pulses, the initial kinetic energy of the weakly photo-doped system is lower and the phase space for scattering is larger. If the photodoping is increased, the density of singlons decreases and the doublons and holons can move only via second order hopping processes. This slows down the thermalization dynamics. In the extreme photo-doping case $d\sim 0.49$, the heating is sufficiently suppressed that an $\eta$ pairing state with a large staggered superconducting order parameter emerges during the pulse \cite{entropy_cooling1}.

\begin{figure}[t]
\includegraphics[width=0.48\textwidth]{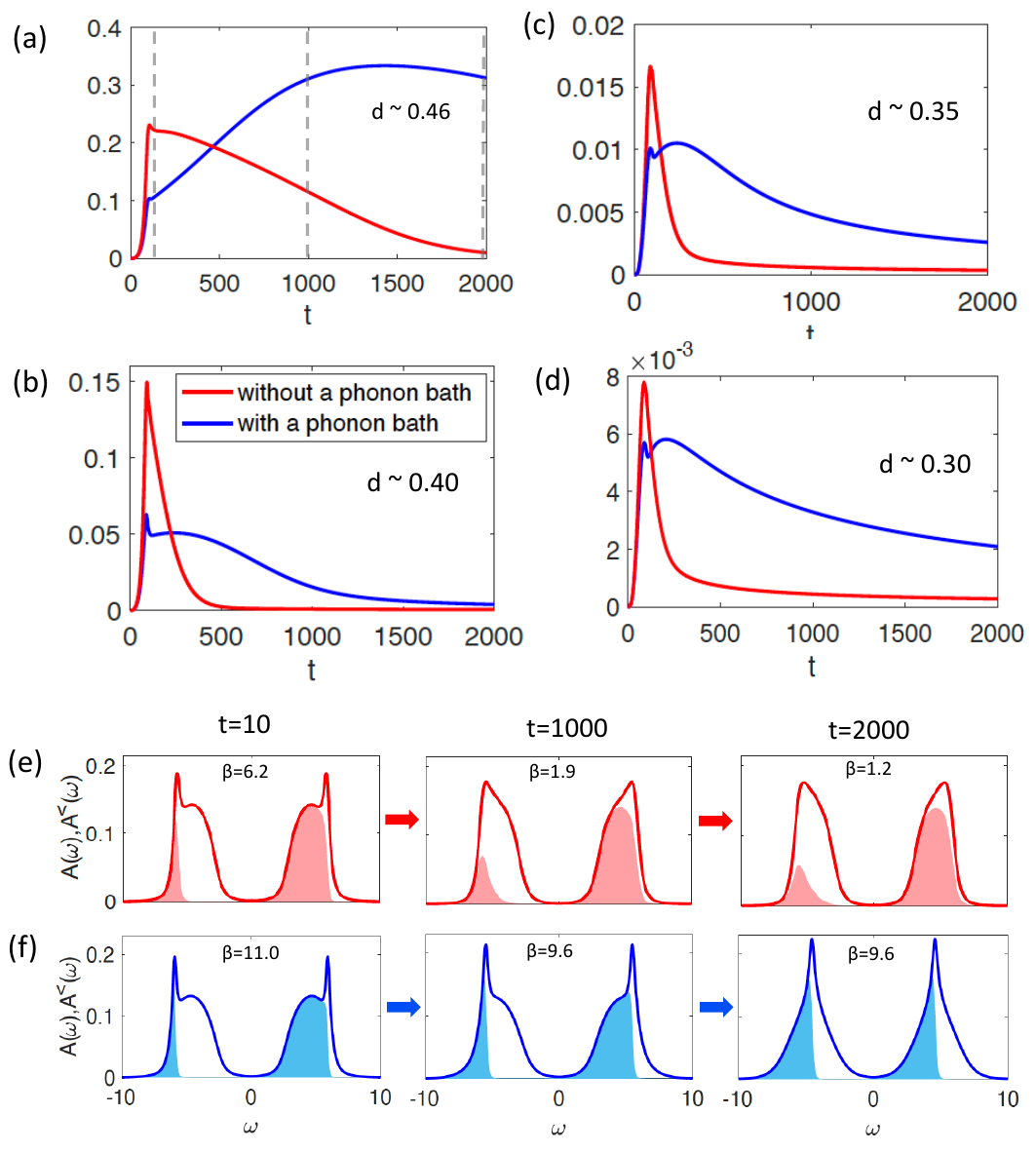}
\caption{(a-d) Evolution of $\eta^{x}$ order with time for 
$U=9$ and the same pulses as in 
\figref{fig2}, with and without phonon coupling. The doublon densities are $d\sim 0.46$ in (a), $d\sim 0.4$ in (b), $d\sim 0.35$ in (c), $d\sim 0.3$ in (d), corresponding to the red, yellow, green and blue curves in ~\figref{fig2}. All these doublon densities are measured just after the laser pulse is switched off at $t=100$. (e,f) Time evolution of the spectral function for $d \sim 0.4$ and $U=9$ at three different times indicated by the gray dashed lines in (a) without a phonon bath (red curves in (e)) and with phonon coupling (blue curves in (f)). The solid line represents the total density of states $A(\omega)$ and the shaded area represents the occupied density of states $A^{<}(\omega)$ with $\beta$ indicating the effective temperatures of the photo-carriers.
}
\label{fig3}
\end{figure}

Next, we focus in the effect of the phonon coupling on the dynamics of the photodoped Mott insulators, which is shown in ~\figref{fig2}(c) and (d). The boson bath allows the system to cool down by dissipating the excess kinetic energy of the photocarriers and absorbing the energy released by doublon-holon recombination processes. At long times, these systems are expected to thermalize at the temperature of the boson bath (here $\beta_\text{boson}=30.0$), while at short times, the dissipation leads to a relaxation within the Hubbard bands. In the simulations, we weakly couple the Mott system to phonons with frequency $\omega_{0}=0.2$, which is much smaller than the Mott gap. As shown in ~\figref{fig2}(c), the doublon density now decays after the pulse and eventually drops below $d=0.25$, which is a prerequisite for thermalization in a positive temperature state. As the photo-doping increases the decay becomes slower, which again can be attributed to the lack of first order hoppings of doublons and holons in the absence of singlons (note that the phonons in Eq.~(\ref{eq_12}) couple to density fluctuations on a given site). The intra-Hubbard band cooling process leads to the initial decrease in the kinetic energy shown in \figref{fig2}(d). For the lowest photo-doping we observe an increase in kinetic energy at $t\gtrsim 1000$, which indicates a reheating of the system through recombination processes facilitated by the boson bath. At later times, once most of the excess doublons and holons have disappeared, the kinetic energy will reach the thermal value indicated by the arrow. It is clear from the figure that this full thermalization occurs on timescales which are much longer than those reached in the simulations.

Next we study the evolution of the $\eta$-SC order for different photo-dopings with and without a phonon bath. Figure~\ref{fig3}(a-d) shows $\eta^{x}$ as a function of time for different doublon densities $d$ (the indicated values are measured after the laser pulse is switched off at $t=100$). We observe that for $d\sim 0.46$, without a phonon bath, the $\eta^{x}$ order grows to $\sim 0.22$ while the pulse is applied, and then decays slowly because of the slow heating effect. If the system is coupled to a phonon bath, $\eta^{x}$ increases less during the pulse, but then grows from $0.1$ to $0.3$ because of the phonon cooling mechanism. For lower photodoping the $\eta^{x}$ order initially stabilizes and then decays. Although the decay rate of the $\eta^{x}$ order is reduced by the phonon coupling, this is not enough to sustain an $\eta$-SC order over several thousand inverse hopping times ($\sim$ several ps in realistic systems) for $d \sim 0.4$, as shown in ~\figref{fig3}(b). For even lower photodopings, the $\eta$-SC order never grows to high values (see ~\figref{fig3}(c,d)), which is consitent with previous studies. Irrespective of doping we observe that a phonon coupling 
either stabilizes the $\eta^{x}$ order (for $d \sim 0.46$) or reduces its decay rate (for $d \lesssim 0.4$).  In ~\figref{fig3}(e) and (f), we show the evolution of the spectral function $A(\omega)$ (total density of states) and $A^{<}(\omega)$ (occupied density of states) for $d\sim 0.4$ at three different times, $t=100$ (just after switching off the laser pulse), $t=1000$ and $t=2000$, which are indicated by the gray dashed lines in ~\figref{fig3}(a). We see that without phonon coupling, the photo-doped state is initially produced with an effective inverse temperature $\beta_{\text{eff}}=6.2$. At later times the effective temperature increases, which is consistent with the kinetic energy plot in ~\figref{fig2}. At even longer times, we expect thermalization in a negative temperature state. In contrast, if we couple the system to phonons, the heating effect is strongly suppressed and the system remains cold with $\beta_{\text{eff}}=11.0$ immediately after the pulse and $\beta_{\text{eff}}=9.6$ at the subsequent times. In this case, due to the rapid decay of the doublon density (see ~\figref{fig2}(c)), the shape of the spectral function changes with time. The decreasing photo-doping concentration at approximately fixed temperature explains the observed decrease in the $\eta$-SC order. This result also shows that the decrease in the kinetic energy observed in Fig.~\ref{fig2}(d) is not only due to a decrease in the effective temperature of the photo-carriers, but to a large extent due to the recombination of these carriers. 

\subsection{$U$ dependence of the dynamics}

\begin{figure}[t]
\includegraphics[width=0.48\textwidth]{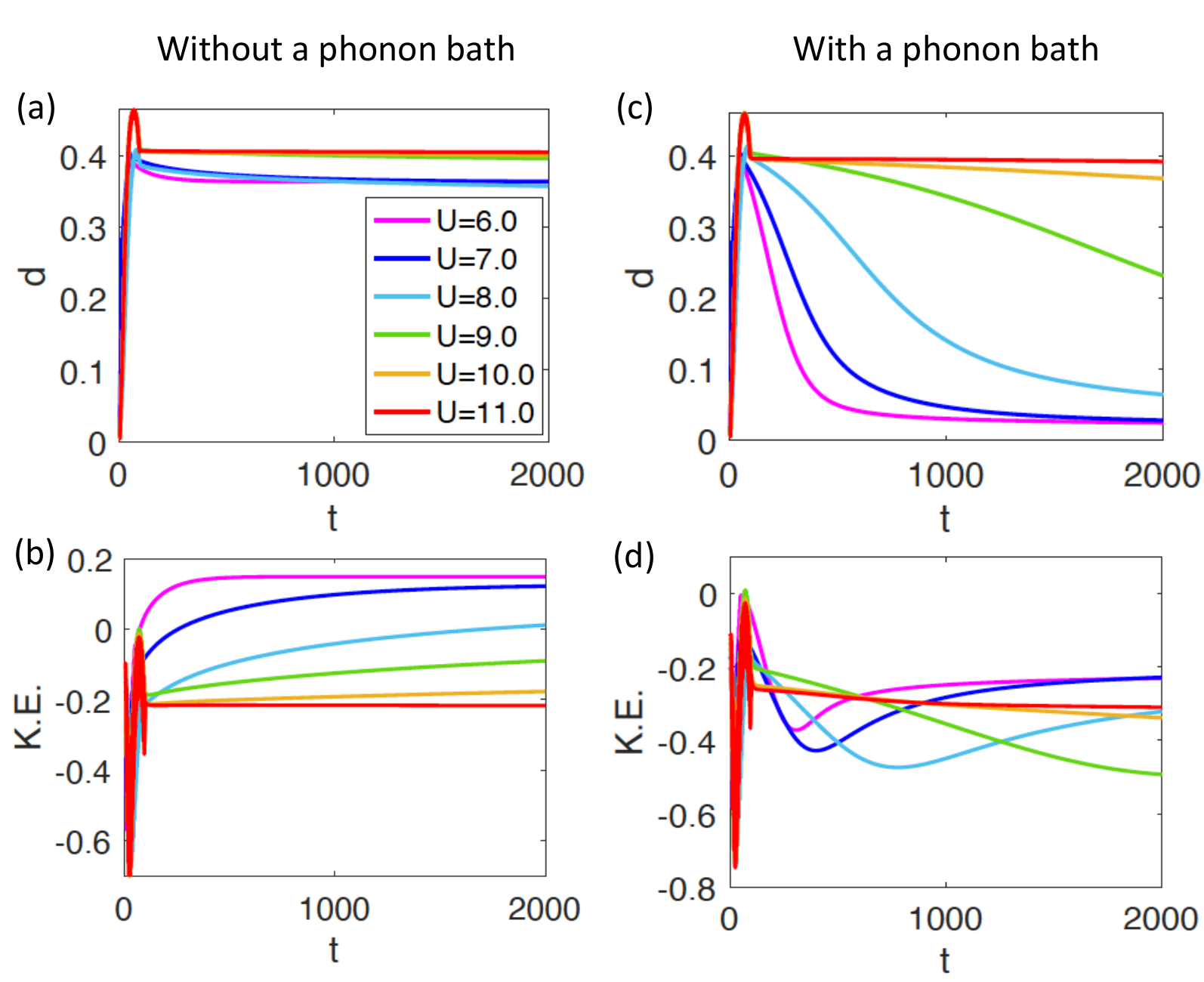}
\caption{(a) Doublon densities and (b) kinetic energies as a function of time without any phonon bath coupling to the system. (c) Doublon densities and (d) kinetic energies as a function of time with a weakly coupled phonon bath. All the results are obtained for roughly the same doublon density $d \sim 0.4$, measured just after the the pulse is switched off at $t=100$. The same photodoping for different $U$ is reached with carefully chosen short chirped pulses. Same colored curves correspond to the same system in all the four figures.}
\label{fig4}
\end{figure}

\begin{figure*}[t]
\includegraphics[width=0.98\textwidth]{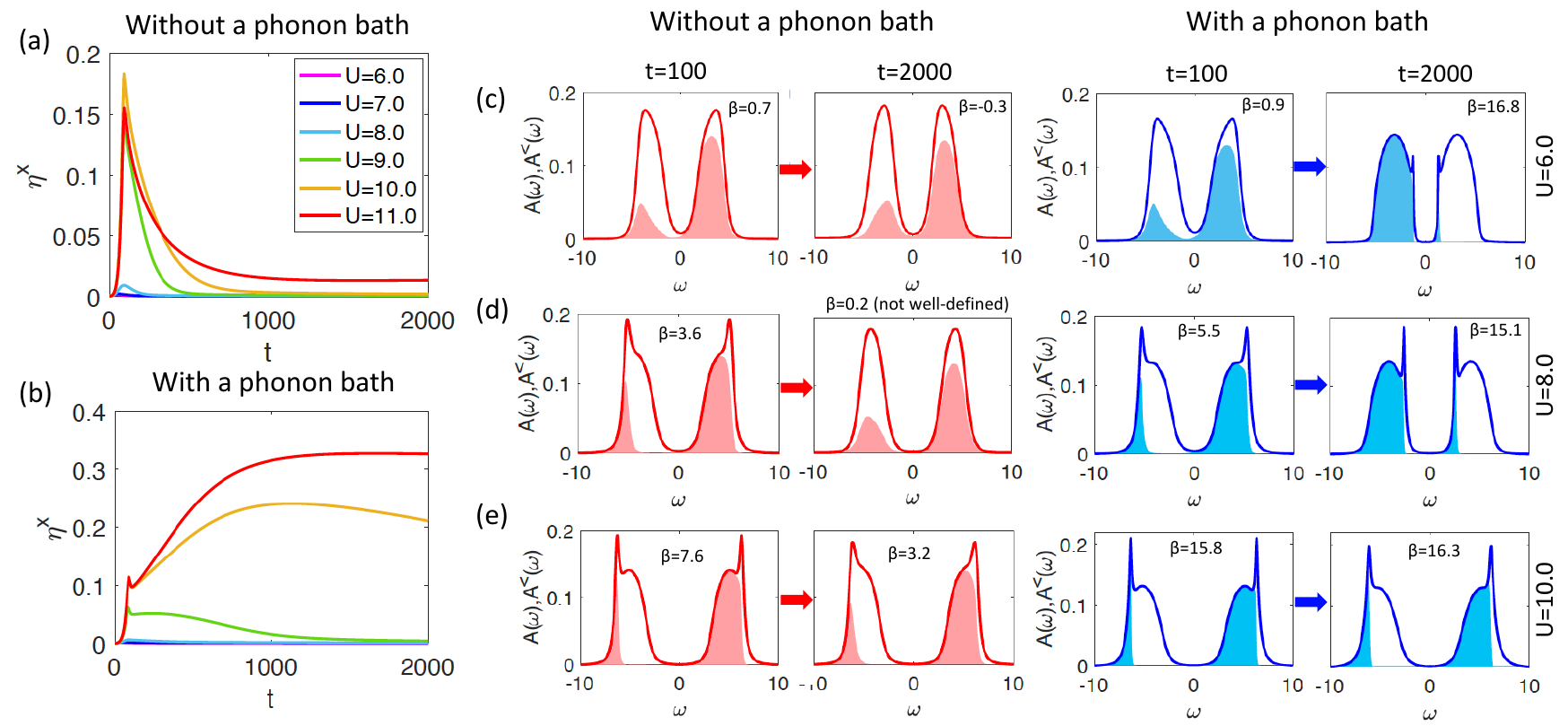}
\caption{(a,b) SC order parameter $\eta^{x}$ as a function of time for different $U$ and roughly the same photodoping $d \sim 0.4$, without a phonon coupling in (a) and with  phonon coupling in (b). (c,d) Time evolution of the spectral function for $d \sim 0.4$ at two different times: after the laser pulse is switched off ($t=100$) and at the longest time of our calculation ($t=2000$). We show results for simulations without a phonon bath (red curves) and with a phonon coupling (blue curves) for $U=6$ (c), $U=8$ (d) and $U=10$ (e). The solid line shows the total density of states $A(\omega)$ and the shaded area represents the occupied density of states $A^{<}(\omega)$ with $\beta$ indicating the effective temperature of the doublons and holons.}
\label{fig5}
\end{figure*}

Next, we study the long-time dynamics of photodoped systems, specifically $\eta$-SC states, for different values of the Hubbard repulsion $U$. It has been pointed out in previous studies that after photodoping the thermalization timescale depends sensitively on the value of $U$. For larger values of $U$, the thermalization process becomes slower due to the lack of multi-particle scattering events capable of converting a potential energy of order $U$ into kinetic energy. For $U\gg W$ ($W$ being the half-bandwidth), the thermalization timescale is given by $\tau \propto \exp{(\alpha U\log(U/W)/W)}$ 
with a numerical constant $\alpha$ of order one
\cite{Sensarma2010}. For $U\gg W$, the doublon density is almost constant on the accessible timescales. Once $U$ is reduced to approximately $W$, the doublon density quickly decays to a thermalized value with an effective temperature ($T_\text{eff}$). Here, we study in detail the effect of this $U$-dependent thermalization timescale on the $\eta$-SC order. We first produce photodoped systems with doublon density $d\sim 0.4$ for different $U$ by using appropriately adjusted chirped pulses as shown in ~\figref{fig4}(a). After the pulse is switched off, we can clearly see the distinct dynamics for $U\lesssim 8$ and $U\gtrsim 8$. At $U=8$, the Mott gap becomes equal to the bandwidth. Panels (a) and (b) show how in the absence of  a phonon bath, the energy injected by the pulse is converted into kinetic energy. 
For $U=6$, this conversion is fast and the kinetic energy increases quickly. After a short time the kinetic energy for $U=6$ becomes positive, which indicates a negative temperature. Indeed, in the absence of phonon coupling, the photo-doped state reaches a negative effective temperature (see~\figref{fig5} (c)). 
Again, the situation becomes different if the system is coupled to a phonon bath, which provides a dissipation channel. In this case, the injected energy need not be converted into kinetic energy and this facilitates the doublon-holon recombination process, as shown in ~\figref{fig5}(c). We observe that for $U<8$, the 
doublon density decays 
to much smaller values compared to the situation without a phonon bath. For $U>8$, the decay process is slower, due to the mentioned thermalization bottleneck. For the largest $U$, the kinetic energy of the system decreases, because of the intra-band cooling of the photo-carries.  This is opposite to the result without phonon coupling. Indeed, fits to the distribution functions confirm the intra-band cooling of the charge carriers
(see ~\figref{fig5} (d),(e)).

Next, we study the effect of such heating and cooling processes on the evolution of the $\eta$-SC order. In \figref{fig5}(a,b), we show the $\eta^{x}$-SC order as a fuction of time for several $U$ with and without a phonon coupling. As mentioned earlier, when there is no phonon coupling, the system heats up quickly after the pulse is switched off. This can be clearly seen in the spectral function plots in \figref{fig5}(c,d,e). Here we plot the spectral functions at $t=100$ (just after the pulse has been switched off) and at the much later time $t=2000$. For $U=6$, the system is initially prepared at $\beta_\text{eff}=0.7$ and it evolves into a negative temperature state ($\beta_\text{eff}=-0.3$) at $t=2000$. For such a high $T_\text{eff} (=1/\beta_\text{eff})$, the $\eta^{x}$-SC order cannot grow and remains very low (\figref{fig5}(a)). As we increase $U$, a lower $T_\text{eff}$ can be reached in the initially prepared state, namely $\beta_\text{eff}=3.6$ for $U=8$ (\figref{fig5}(d)) and $\beta_\text{eff}=7.6$ for $U=10$ (\figref{fig5}(e)). As a result, initially the $\eta^{x}$ order grows to a high value ($\sim 0.15$ for $U \ge 9$) as can be seen from \figref{fig5}(a). However, the dynamics after the end of the pulse depends on the value of $U$. For larger $U$, the dynamics becomes slower. Around $U=8$, the system transitions from a positive to a negative temperature state at $t=2000$ and $\beta_\text{eff}$ is not well-defined due to a distribution function with a distinctly non-Fermi like shape within the Hubbard bands. 
At $U=10$, the system still has a high (positive) effective temperature  ($\beta_{\text{eff}}=3.2$) at $t=2000$. If we could run the simulations up to even longer times, eventually all the systems with high photodoping would evolve to a negative temperature state. 
As a result of this heating process, $\eta^{x}$ starts to decay after the pulse. As expected, the decay process is slower for higher $U$ (\figref{fig5}(a)) due to the slower dynamics.

To stabilize the $\eta$-SC state, we need to suppress the heating, which can be achieved by coupling the system to a phonon bath. A strong enough phonon coupling prevents the system from switching into a (transient) negative temperature state. The primary effects of the phonon coupling are twofold -- for large enough $U$, it helps to decrease the effective temperature of the photo-carriers, while for smaller $U$, it leads to rapid doublon-holon recombination. The photo-doped states with $U=6$ and $8$, initially prepared with $\beta_{\text{eff}}=0.9$ and $5.5$, respectively, evolve to states with a low doublon density ($d<0.1$) and with low effective temperature ($\beta_{\text{eff}}=16.8$ and $15.1$, respectively). Despite these low effective temperatures, these systems do not show a significant $\eta^{x}$ order because of the rapid decay of the doublons. A large $\eta^{x}$ order can only be sustained in a system with both large enough $U$ and with a phonon coupling. In our calculations, this is achieved for $U\gtrsim 10$. In the blue panels of \figref{fig5}(e), we can see that the initially prepared state at $\beta_{\text{eff}}=15.8$ remains qualitatively unchanged and evolves to a colder temperature state ($\beta_{\text{eff}}=16.3$) with almost the same photo-carrier density at $t=2000$. As a result, the $\eta^{x}$ order grows from $0.1$ to $0.2$ and saturates to an almost steady value around $t=1000$ (\figref{fig5}(b)). For $U=11$, $\eta^{x}$ is even more stable with a value above $0.3$. Such a plateau in the dynamics of the $\eta^{x}$ order along with an almost constant low effective temperature suggest the relaxation of the system into a long-lived prethermalized steady state.

\subsection{Effect fo $\eta$-SC order on the relaxation dynamics}
\begin{figure}[t]
\includegraphics[width=0.5\textwidth]{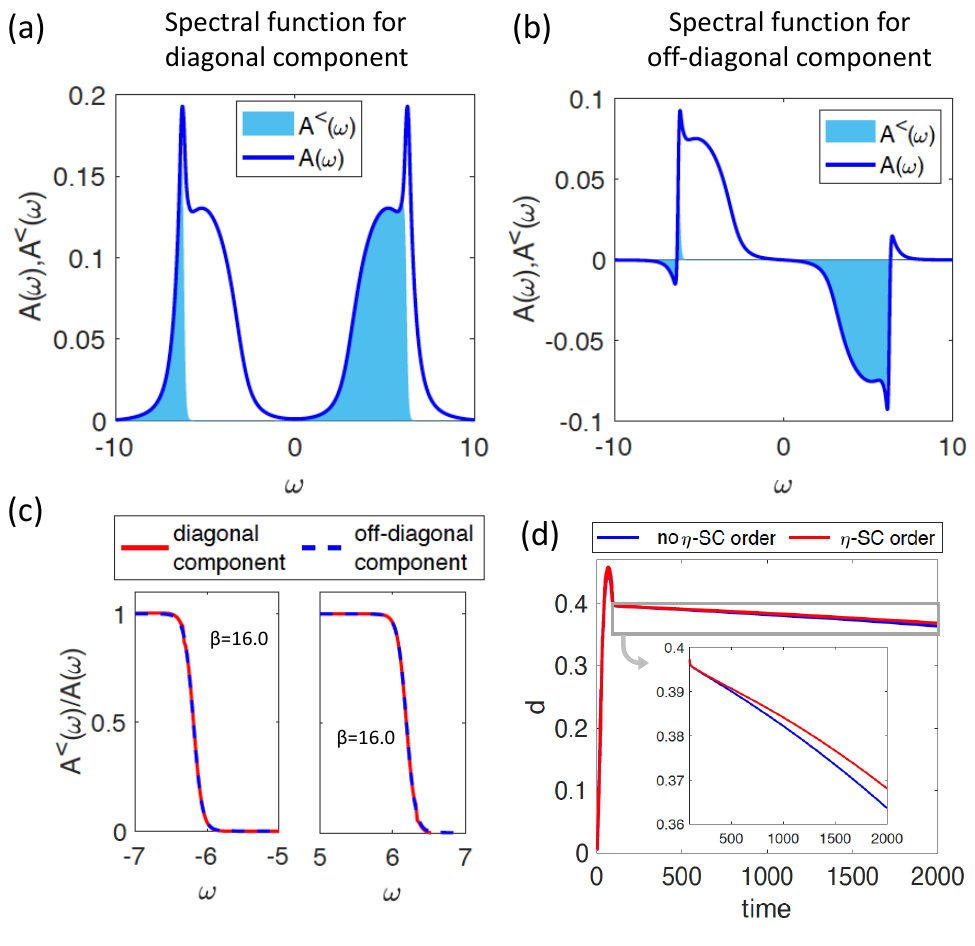}
\caption{Total density of states $A(\omega)$ and occupied density of states $A^{<}(\omega)$ for (a) the diagonal component and (b) off-diagonal anomalous component of the Green's function obtained from real-time simulation after the laser pulse is switched off at $t=100$. (c) The ratio between $A^{<}(\omega)$ and $A(\omega)$ for both components, which resembles the Fermi distribution function. The effective temperature extracted from Fermi function fits is $\beta=16.0$ for both components. (d) Doublon density as a function of time for a system with $\eta$-SC order and without $\eta$-SC order (the inset shows the zoomed version indicated by the gray rectangle). The calculations are for $U=10$ and $d=0.397$.
}
\label{fig8}
\end{figure}
Here we investigate the evolution of $\eta$-SC order and its effect on the relaxation dynamics of the photodoped system. We study the system with $U=10$ and doublon density $d\sim 0.4$. Figure~\ref{fig8}(a) shows the spectral function of the diagonal component of the Green's function at $t=1000$, when the photodoped state has a high $\eta$-SC order $\eta^{x}\sim 0.22$ (see ~\figref{fig5}(b)). We also plot the spectral function of the anomalous off-diagonal component of the Green's function which is shown in ~\figref{fig8}(b). We extract the effective temperature for both components in the upper and lower Hubbard bands by fitting a Fermi distribution function to the ratio of $A^{<}(\omega)$ and $A(\omega)$. The effective temperatures are the same, $\beta = 16.0$ (see \figref{fig8}(c)). This indicates that the system evolves with a well defined effective temperature, which can be reproduced by the steady state NESS simulation. 

We also considered the effect of the $\eta$-SC order on the doublon decay during the relaxation process. In ~\figref{fig8}(d) we plot the doublon density as a function of time for two systems with $U=10, d\sim 0.4$ -- one with $\eta$-SC order and one without $\eta$-SC order. It turns out that the doublon decay rate is almost the same. Only a small deviation is observed at longer times which suggests that the $\eta$-SC state provides a little more stability to the photodoped system and reduces the doublon decay rate by a small amount.

\subsection{Comparison to nonequilibrium steady states} 
An interesting question is whether the long-lived prethermal states that emerge in the time evolution of the system can be described by a quasi-steady state formalism (NESS approach).
If we focus on the doublon densities, then without any phonon coupling these densities approach a thermalized value on the accessible timescales for small $U$, but remain almost constant for large $U$. However, thermalization in a negative-temperature state or a very slow heating of the electronic state after a laser pulse is not realistic in a solid state system. There, we expect that the electrons couple to bosonic modes, e.g. phonons, and hence dissipate energy. In the more realistic simulations with phonon coupling, the charge carriers can relax within the Hubbard bands (dissipate their kinetic energy) while for sufficiently large gap, the recombination of charge carriers is slow. The system then reaches a quasi-steady state with an almost constant spectral function and a well-defined effective temperature of the photo-doped charge carriers.

\begin{figure}[t]
\includegraphics[width=0.48\textwidth]{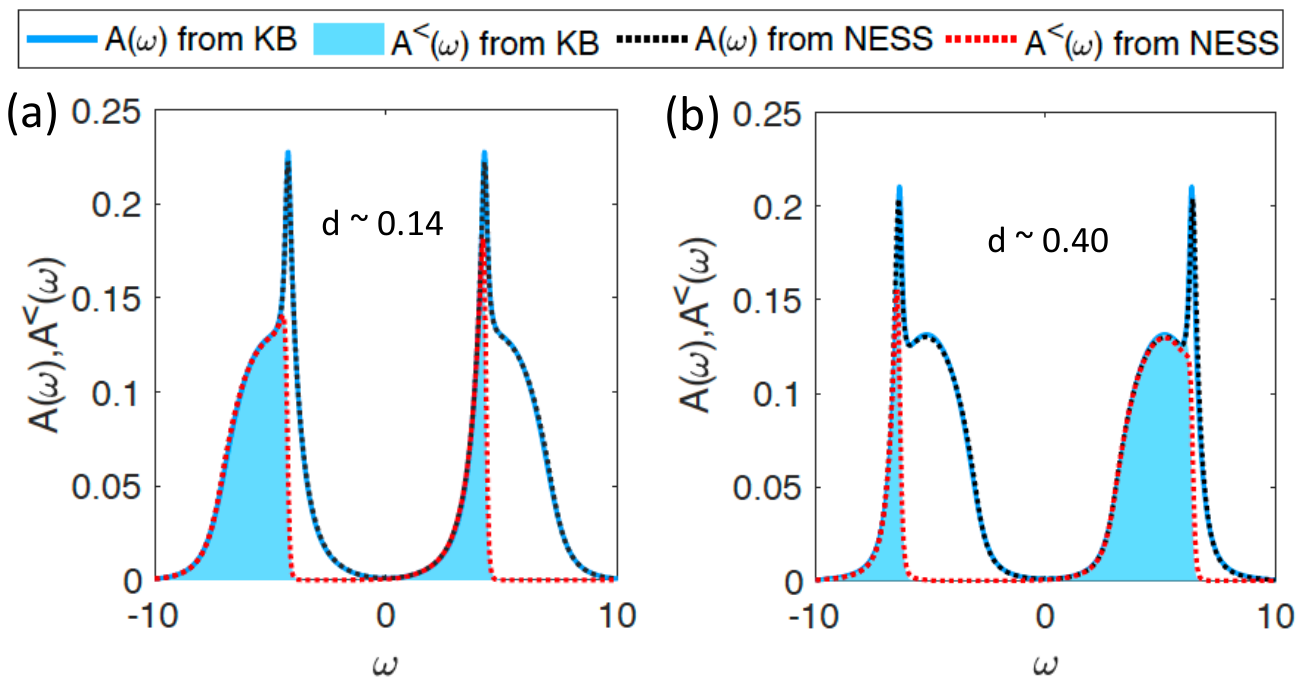}
\caption{Total density of states $A(\omega)$ and occupied density of states $A^{<}(\omega)$ obtained from the real-time simulation on the Kadanoff-Baym (KB) contour (blue curve and blue shaded region) and from the NESS simulation (black and red dashed curve) with suitably adjusted bath parameters.
}
\label{fig7}
\end{figure}

To prepare true steady state solutions, we use the NESS formalism, where two cold non-interacting Fermion baths with half-bandwidth $W_{b}=2.0$ are placed at $\pm U/2$ and the system is coupled to phonon baths with the phonon frequency $\omega_{0}=0.4$ and phonon coupling strength $g=0.2$. By adjusting the coupling to the Fermion bath $\Gamma$ and the temperature of the Fermion baths $T_{\text{b}}$, we are able to create photodoped steady states with an effective temperature very close to that obtained in the quasi-steady state of the real-time simulations. 
Figure~\ref{fig7} compares these spectral functions for photodoped states with $U=10.0$ and the two doublon densities $d=0.14$ and $0.4$ at inverse effective temperature $\beta_{\text{eff}}\sim16.0$. The spectral functions from the real-time simulation is calculated after the laser pulse is switched off at $t=100$. We find a close correspondence between the NESS steady state result and the real-time simulation result for both $A(\omega)$ and $A^{<}(\omega)$, as shown in \figref{fig7}. The agreement is particularly good for the lower photodoping shown (\figref{fig7}(a) for $d=0.14$), while the deviation between the two spectra is a bit larger for 
$d\sim0.4$,  
as shown in \figref{fig7}(b). This is related to the fact that for larger photodoping, a stronger coupling to the cold Fermion baths is needed in the NESS approach. These couplings have a small but nonnegligible effect on the spectral function (beyond imposing the effective temperature of the photocarriers). But overall, even for  photodoping as large as $d=0.4$, the steady state calculation provides a good description of the $\eta$-SC state solutions as obtained from the real-time simulation with phonon dissipation. Thus, our results show that both the real-time and NESS approaches yield consistent results for large gap phototdoped Mott insulators, and that the NESS approach can be used to study the long-lived prethermalized $\eta$-SC orders in these systems.

\section{Conclusions}
We studied the long-time dynamics of photodoped Mott insulators both without and with a phonon coupling and showed that it is possible to induce an $\eta$-SC state with an ultrashort laser pulse. The evolution of the $\eta$-SC order depends on the relaxation dynamics which follows after the pulse. In the absence of an energy dissipation mechanism, the $\eta$-SC order quickly decays, especially in small-gap systems, due to the heating effect. By coupling a system with sufficiently large gap to phonon baths, it is possible to transiently stabilize or even enhance the $\eta$-SC order. The presence of the off-diagonal order does not have a significant effect on the life-time of the photo-doped state, and the effective temperatures extracted from the normal and anomalous distribution functions are consistent. This shows that with phonon coupling, a long-lived prethermalized state with well defined effective temperature of the doublons and holons can be sustained for more than $1000$ inverse hopping times. Such simulation times are accessible using  memory-truncated Kadanoff Baym equations.

Our study furthermore demonstrated that the quasi-steady states realized in real-time simulations with phonon coupling are well described by the NESS approach, and it clarified the regime (sufficiently large gap, not too strong photo-doping) where this effective description is valid. Interactions with phonons and other bosonic degrees of freedom like spin- or Hund's excitations are present in realistic materials and typically result in intra-Hubbard band relaxation on a sub-ps timescale. The NESS formalism enables a good control over the density and effective temperature of the photo-carriers in the resulting long-lived prethermalized state. Since the complexity of NESS simulations is lower than that of real-time simulations, it is a promising platform for the development of more accurate and efficient numerical techniques \cite{Thoenniss2023,Mrtin_inchworm,Eckstein2024,Kim2024}, which in the future will enable a more quantitative description of photo-excited Mott states. This will be helpful in the search of new types of hidden orders in (multi-orbital) Mott systems. 
\label{sec_conclusion}

\acknowledgements

The calculations were run on the beo06 cluster at the University of Fribourg. S.R. and P.W. acknowledge support from 
SNSF Grant No. 200021-196966.
M.E. acknowledges funding through the DFG QUAST-FOR$5249$ - $449872909$ (Project P$6$), and through the Cluster of Excellence ``CUI: Advanced Imaging of Matter“ of the DFG – EXC $2056$ – project ID $390715994$.


\begin{thebibliography}{99}

\bibitem{Cavalleri2018} A. Cavalleri, Photo-induced superconductivity, Contemp. Phys. {\bf 59}, 31 (2018).

\bibitem{Fausti2011} D. Fausti, R. I. Tobey, N. Dean, S. Kaiser, A. Dienst, M. C. Hoffmann, S. Pyon, T. Takayama, H. Takagi and A. Cavalleri, Light-induced superconductivity in a stripe-ordered cuprate, Science {\bf 331}, 189 (2011).

\bibitem{phsc_FeSe}T. Suzuki, T. Someya, T. Hashimoto, S. Michimae, M. Watanabe, M. Fujisawa, T. Kanai, N. Ishii, J. Itatani, S. Kasahara, Y. Matsuda, T. Shibauchi, K. Okazaki and S. Shin, Photoinduced possible superconducting state with long-lived disproportionate band filling in FeSe. Commun. Phys. {\bf 2}, 115 (2019).

\bibitem{Buzzi_2020}M. Buzzi, D. Nicoletti, M. Fechner, N. Tancogne-Dejean, M. A. Sentef, A. Georges, T. Biesner, E. Uykur, and M. Dressel {\it et al}., Photomolecular High-Temperature Superconductivity, Phys. Rev. X {\bf 10}, 031028 (2020).

\bibitem{phind_sc} M. Mitrano, A. Cantaluppi, D. Nicoletti, S. Kaiser, A. Perucchi, S. Lupi, P. Di Pietro, D. Pontiroli, M. Ricco, S. R. Clark, et al., Possible light-induced superconductivity in K$_3$C$_{60}$ at high temperature, Nature {\bf 530}, 461 (2016).

\bibitem{phind_sc1} M. Budden, T. Gebert, M. Buzzi, G. Jotzu, E. Wang, T. Matsuyama, G. Meier, Y. Laplace, D. Pontiroli, M. Ricco, F. Schlawin, D. Jaksch, and A. Cavalleri, Evidence for metastable photo-induced superconductivity in K$_3$C$_{60}$, Nat. Phys. {\bf 17}, 611 (2021).

\bibitem{Murakami_rev}Y. Murakami, D. Golez, M. Eckstein, P.Werner, Photo-induced nonequilibrium states in Mott insulators, arXiv:2310.05201.

\bibitem{Strohmaier2010} N. Strohmaier, D. Greif, R. J\"ordens, L. Tarruell, H. Moritz, T. Esslinger, R. Sensarma, D. Pekker, E. Altman and E. Demler, Observation of Elastic Doublon Decay in the Fermi-Hubbard Model, Phys. Rev. Lett. {\bf 104}, 080401 (2010).

\bibitem{Sensarma2010} R. Sensarma, D. Pekker, E. Altman, E. Demler, N. Strohmaier, D. Greif, R. J\"ordens, L. Tarruell, H. Moritz and T. Esslinger, Lifetime of double occupancies in the Fermi-Hubbard model, Phys. Rev. B {\bf 82}, 224302 (2010).

\bibitem{Eckstein2011} M. Eckstein and P. Werner, Thermalization of a pump-excited Mott insulator, Phys. Rev. B {\bf 84}, 035122 (2011).

\bibitem{Lenarcic2013} Z. Lenarčič and P. Prelovšek, Ultrafast charge recombination in a photoexcited mott-hubbard insulator. Phys. Rev. Lett. {\bf 111}, 016401 (2013).

\bibitem{Lenarcic2014} Z. Lenarcic and P. Prelovsec, Charge recombination in undoped cuprates, Phys. Rev. B {\bf 90}, 235136 (2014).

\bibitem{yuta_etasc} Y. Murakami, S. Takayoshi, T. Kaneko, A. M. L\"auchli, and P. Werner, Spin, Charge, and $\eta$-Spin Separation in One-Dimensional Photodoped Mott Insulators, Phys. Rev. Lett. {\bf 130}, 106501 (2023).

\bibitem{yuta_etasc1} Y. Murakami, S. Takayoshi, T. Kaneko, Z. Sun, D. GoleÅŸ, A. J. Millis, and P. Werner, Exploring nonequilibrium phases of photo-doped Mott insulators with generalized Gibbs ensembles, Comm. Phys. {\bf 5}, 23 (2022).

\bibitem{eta_jiajun} J. Li, D. Golez, P. Werner, and M. Eckstein, $\eta$-paired superconducting hidden phase in photodoped Mott insulators, Phys. Rev. B {\bf 102}, 165136 (2020).

\bibitem{eta_spintrip} S. Ray, Y. Murakami, and P. Werner, Nonthermal superconductivity in photodoped multiorbital Hubbard systems, Phys. Rev. B {\bf 108}, 174515 (2023).

\bibitem{eta_spintrip1} S. Ray and P. Werner, Photoinduced ferromagnetic and superconducting orders in multiorbital Hubbard models, Phys. Rev. B {\bf 110}, L041109 (2024).

\bibitem{kaneko1} S. Ejima, T. Kaneko, F. Lange, S. Yunoki, and H. Fehske, Photoinduced $\eta$-pairing at finite temperatures, Phys. Rev. Research {\bf 2}, 032008(R) (2020).

\bibitem{kaneko2} T. Kaneko, S. Yunoki, and A. J. Millis, Charge stiffness and long-range correlation in the optically induced $\eta$-pairing state of the one-dimensional Hubbard model, Phys. Rev. Research {\bf 2}, 032027(R) (2020).

\bibitem{tri_chiral} J. Li, M. M\"uller, A. Kim, A. L\"aeuchli, and P. Werner, Twisted chiral superconductivity in photodoped frustrated Mott insulators, Phys. Rev. B {\bf 107}, 205115 (2023).

\bibitem{NESSi} M. Sch\"uler, D. Golelez, Y. Murakami, N. Bittner, A. Herrmann, H. U. R. Strand, P. Werner, and M. Eckstein, NESSi: The Non-Equilibrium Systems Simulation package, Comput. Phys. Commun. {\bf 257}, 107484 (2020).

\bibitem{entropy_cooling} P. Werner, M. Eckstein, M. M\"uller, and G. Refael, Light-induced evaporative cooling of holes in the Hubbard model, Nat. Comm. {\bf 10}, 5556 (2019).

\bibitem{entropy_cooling1} P. Werner, J. Li, D. Golez, and M. Eckstein, Entropy-cooled nonequilibrium states of the Hubbard model, Phys. Rev. B {\bf 100}, 155130 (2019).

\bibitem{NESS_martin} J. Li and M. Eckstein, Nonequilibrium steady-state theory of photodoped Mott insulators, Phys. Rev. B {\bf 103}, 045133 (2021).

\bibitem{Mrtin_inchworm} F. K\"unzel, A. Erpenbeck, D. Werner, E. Arrigoni, E. Gull, G. Cohen, and M. Eckstein, Numerically Exact Simulation of Photodoped Mott Insulators, Phys. Rev. Lett. {\bf 132}, 176501 (2024).

\bibitem{Schueler2018} M. Schüler, M. Eckstein, and P. Werner, Truncating the memory time in nonequilibrium dynamical mean field theory calculations, Phys. Rev. B {\bf 97}, 245129 (2018).

\bibitem{trunc_NESSi} C. Stahl, N. Dasari, J. Li, A. Picano, P. Werner, and M. Eckstein, Memory truncated Kadanoff-Baym equations, Phys. Rev. B {\bf 105}, 115146 (2022).

\bibitem{Georges1996} A. Georges, G. Kotliar, W. Krauth, and M. J. Rozenberg, Dynamical mean-field theory of strongly correlated fermion systems and the limit of infinite dimensions, Rev. Mod. Phys. {\bf 68}, 13 (1996).

\bibitem{Eckstein2010} M. Eckstein and P. Werner, Nonequilibrium dynamical mean-field calculations based on the noncrossing approximation and its generalizations, Phys. Rev. B {\bf 82}, 115115 (2010).

\bibitem{Martin_phonon} M. Eckstein and P. Werner, Photoinduced States in a Mott Insulator, Phys. Rev. Lett. {\bf 110}, 126401 (2013).

\bibitem{Thoenniss2023} J. Thoenniss, M. Sonner, A. Lerose, and D. A. Abanin, Efficient method for quantum impurity problems out of equilibrium, Phys. Rev. B {\bf 107}, L201115 (2023). 

\bibitem{Eckstein2024} M. Eckstein, Solving quantum impurity models in the non-equilibrium steady state with tensor trains, arXiv:2410.19707  (2024).

\bibitem{Kim2024} A. Kim and P. Werner, Strong coupling impurity solver based on quantics tensor cross interpolation, arXiv:2411.19026   (2024).

\end{thebibliography}
\end{document}